\documentclass[fleqn,12pt,twoside]{article}
\usepackage{espcrc1}
\usepackage{graphicx}

\newcommand{\AmS}{{\protect\the\textfont2
  A\kern-.1667em\lower.5ex\hbox{M}\kern-.125emS}}
\title{Radii of weakly bound three-body systems: halo nuclei 
and molecules}

\author{M.T. Yamashita
\address{Laborat\'orio do Acelerador Linear, Instituto de
F\'{\i}sica, Universidade de S\~{a}o Paulo, \\ 
C.P. 66318, CEP 05315-970, S\~{a}o Paulo, Brazil}, 
T. Frederico
\address{Departamento de F\'\i sica, Instituto Tecnol\'ogico de
Aeron\'autica, CTA, \\
C.P. 12228-900, S\~ao Jos\'e dos Campos, Brazil}, 
R.S. Marques de Carvalho
\address{Instituto de F\'\i sica Te\'orica, Universidade
Estadual Paulista, \\ C.P. 01405-900, S\~{a}o Paulo, Brazil} and
Lauro Tomio$^c$}
       
\begin{document}

\maketitle

\begin{abstract}
A renormalized three-body model with zero-range potential is
used to estimate the mean-square radii of three-body halo nuclei
and molecular systems. The halo nuclei ($^6$He, $^{11}$Li, $^{14}$Be 
and $^{20}$C) are described as point-like inert cores and two neutrons.
The molecular systems, with two helium atoms, are of the type 
$^4$He$_2-$X, where X$=^4$He, $^6$Li, $^7$Li, or $^{23}$Na.
The estimations are compared with experimental data
and realistic results.  
\end{abstract}

\vspace {0.7cm}
Halo nuclei and weakly bound molecules are quantum systems with very 
large sizes in which the constituent particles have a large probability 
to be found much beyond the interaction range. Under this circumstances,
the physical properties of such bound systems can be defined by few physical
scales. 

In the present contribution we report results obtained for 
the root-mean-square radii of three-body halo nuclei and 
molecular systems that are obtained from a {\it universal scaling
function}  calculated within a renormalized scheme for three particles 
interacting through pairwise Dirac-$\delta$ potential. 
In the case where we have the two-body scattering length, $a$,
much greater than the effective range of the potential, $r_0$,
($a/r_0\gg 1$) our zero-range approach is expected to be a
good aproximation~\cite{AdPRA88}. 
In the {\it scaling limit}~\cite{AmPRC97,YaPRA02} all
the observables of the three-body system can be represented by a 
function that depends only on the three- and two-body scales.  

Figure 1 and Table 1 show results for halo nuclei systems, where  
$C$ represents the core and $n$ the neutron. In Fig.~1, our
results ($\hbar=m_n=1$) are given as functions of the mass ratio 
$A\equiv m_C/m_n$, where $m_C$ and $m_n$ are the masses of the core 
and the neutron, respectively. $r^2_\gamma$ ($\gamma\equiv C,\; n$) is the 
mean-square distance from the particle $\gamma$ to the center-of-mass of the system.
$r^2_{n\gamma}$ is the mean-square distance between the particles $n$ and $\gamma$. 
The results given in Fig.~1 are obtained in the limit where the two-body energies are
equal to zero, such that the system have just the three-body energy
as a physical scale. 
Our formalism is presented in detail in ref.~\cite{YaPRA03}, where we show
that Fig. 1 applies equally well to weakly bound molecular systems, as the
results are given in terms of dimensionless quantities.

%%%%%%%%%%%%%%%%%%%%%%%%%%%%%% FIG. 1 %%%%%%%%%%%%%%%%%%%%%%%%%%%%%
\begin{figure}[thbp]
\centerline{\includegraphics*[width=8.5cm,angle=0]{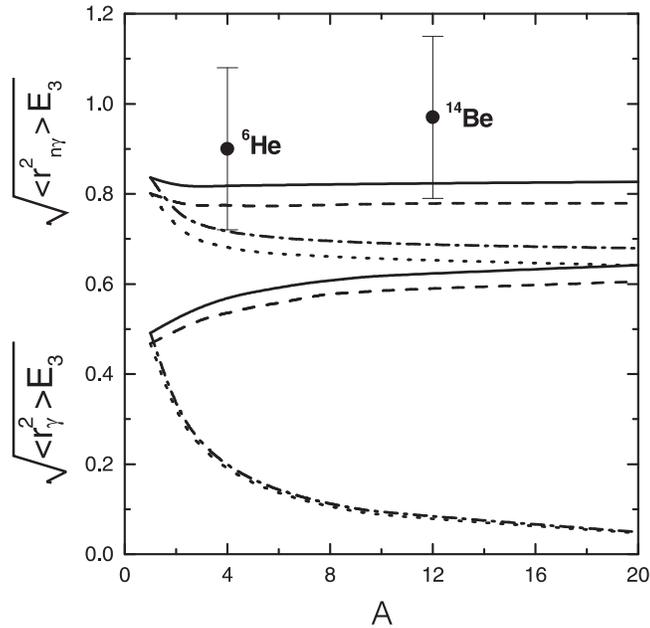}}
\caption[dummy0]
{Dimensionless products $\sqrt{\langle r^2_{n\gamma}\rangle E_3}$ 
[upper (a) plots] and $\sqrt{\langle r^2_{\gamma}\rangle E_3}$ 
[lower (b) plots], are given as functions of $A\equiv m_C/m_n$, 
for $E_{nn}=E_{nC}=0$.
The results for the ground-state are shown with solid line
($\gamma=n$) and dot-dashed line ($\gamma=C$);
and, for the first excited state, with dashed line ($\gamma=n$)
and dotted line ($\gamma=C$).
}
\label{fig1}
\end{figure}
%%%%%%%%%%%%%%%%%%%%%%%%%%%%%%%%%%%%%%%%%%%%%%%%%%%%%%%%%%%%%%%%%%%
\begin{table}[htb]
\caption{Results of the neutron-neutron root-mean-square radii in halo nuclei. 
The core ($C$) is given in the first column, 
the three-body ground state energy and the corresponding 
two-body energy are respectively given in the second and third columns. 
The virtual states are indicated by (v), and the $nn$ virtual state energy is
taken as 143 keV.
The experimental values, in the last column, come from 
ref.~\cite{MaPL00}.}
\label{t1}
\newcommand{\m}{\hphantom{$-$}}
\newcommand{\cc}[1]{\multicolumn{1}{c}{#1}}
\renewcommand{\tabcolsep}{2pc} 
\renewcommand{\arraystretch}{1.2} 
\begin{tabular}{@{}lllll}
\hline
Core & \cc{$E_3^{(0)}$ (MeV)} & \cc{$E_{nC}$ (MeV)} & \cc{$\sqrt{\langle
r_{nn}^2\rangle}$ (fm)} 
& \cc{$\sqrt{\langle r_{nn}^{2\{exp\}}\rangle}$ (fm)} \\
\hline
$^4He$ & \m0.97\cite{AuNPA93} & \m0\cite{MaNPA01} & \m5.1 & \m5.9$\pm$1.2 \\
$^9Li$  & \m0.32\cite{DaPRC94} & \m0.8\cite{DaPRC94}(v) & \m5.9 &
\m6.6$\pm$1.5 \\
$^{12}Be$ & \m1.34\cite{AuNPA93}  & \m0.002\cite{MaPRC97}(v)  & \m4.4  & \m5.4$\pm$1.0  \\
$^{18}C$ & \m3.51\cite{AuNPA93} & \m0.16\cite{AuNPA93} & \m3.0 & \m- \\
\hline
\end{tabular}
\end{table}
%%%%%%%%%%%%%%%%%%%%%%%%%%%%%%%%%%%%%%%%%%%%%%%%%%%%%%%%%%%%%%%%%%%

\begin{table}
\caption[dummy0] {Results for different radii of the molecular
systems $\alpha\alpha\beta$, where $\alpha\equiv ^4$He and   
$\beta$ is identified in the first column. The ground-state
energies of the triatomic molecules and the corresponding energies
of the diatomic subsystems, obtained from ref.~\cite{Yuan}, are given in
the second, third and forth columns. $\langle
r^2_{\alpha\gamma}\rangle$ is the corresponding mean-square
distance between the particles $\alpha$ and $\gamma$ ($= \alpha,
\beta$). $\langle r^2_{\gamma}\rangle$ is the mean-square distance
of $\gamma$ to the center-of-mass of the system. } 
\newcommand{\m}{\hphantom{$-$}}
\newcommand{\cc}[1]{\multicolumn{1}{c}{#1}}
\renewcommand{\tabcolsep}{1.4pc} 
\renewcommand{\arraystretch}{1.2} 
\begin{tabular}{@{}llllllll}
\hline
$\beta$ & $E_3^{(0)}$ & $E_{\alpha\alpha}$ & $E_{\alpha\beta}$ &
$\sqrt{\langle r^2_{\alpha\alpha}\rangle}$ & $\sqrt{\langle
r^2_{\alpha\beta}\rangle} $  & $\sqrt{\langle r^2_\alpha\rangle}$ &
$\sqrt{\langle r^2_\beta\rangle}$ \\
&(mK)&(mK)&(mK)& (\AA) & (\AA)&(\AA)&(\AA)\\
\hline
$^4$He    & 106.0& 1.31 & 1.31 & 9.45 & 9.45 & 5.55 & 5.55  \\
$^6$Li    & 31.4 & 1.31 & 0.12 &16.91 &16.38 & 10.50& 8.14  \\
$^7$Li    & 45.7 & 1.31 & 2.16 &14.94 &13.88 & 9.34 & 6.31  \\
$^{23}$Na &103.1 & 1.31 &28.98 &11.66 & 9.54 & 8.12 & 1.94  \\
\hline
\end{tabular}
\end{table}
%%%%%%%%%%%%%%%%%%%%%%%%%%%%%%%%%%%%%%%%%%%%%%%%%%%%%%%%%%%%%%%%%%%

We can see in Fig.~\ref{fig1} that, although the ratio between
the energies of the ground and the first excited states are
about few hundreds~\cite{univ}, we have practically no difference between 
our dimensionless curves, supporting the validity of our {\it scaling limit}.  
The experimental results for the halo nuclei
$^6$He ($^4$He+n+n) and $^{14}$Be ($^{12}$Be+n+n) 
are quite consistent with our predictions based on the zero-range
calculations.
Therefore, our assumption that such nuclei could be represented by 
inert cores and two neutrons, interacting through short range forces, 
produces a reasonable description of the average interparticle distances.

In Table~\ref{t1} we present the results of the neutron-neutron root-mean-square
distances, for the halo nuclei C-n-n system 
$^6$He, $^{11}$Li, $^{14}$Be, and $^{20}$C. Our results are compared with
experimental values, given in ref.~\cite{MaPL00}. Although the calculations
are in reasonable agreement with data, they systematically underestimate
the measured values. 
Analogously, we show in Table 2 our results, obtained in ref.~\cite{YaPRA03} 
for weakly-bound three-body molecules. In this case, we present results for 
the root-mean-square distances between the particles and also the root-mean-square 
distances of each particle to the center-of-mass of the system.
For the $^4$He$_3$ trimer, our calculations deviate only about 14\% from the
realistic results obtained in ref.~\cite{bk}. 

In conclusion, the present model for the weakly bound halo nuclei and
molecular systems gives a reasonable description of these systems, 
validated by the comparison with experimental data and realistic results.

We would like to thank the partial support from FAPESP and CNPq of Brazil.

\end{document}